\documentclass[twocolumn,english,showpacs,preprintnumbers,prbr]{revtex4-1}
\usepackage[latin9]{inputenc}
\usepackage{amsmath}
\usepackage{amssymb}
\usepackage{graphicx}
\usepackage{esint}
\usepackage{float}
\usepackage{hyperref}
\hypersetup{colorlinks=true, linkcolor=blue, filecolor=blue, urlcolor=blue, citecolor=blue}

\makeatletter
\usepackage{color}
\@ifundefined{textcolor}{}{%
 \definecolor{BLACK}{gray}{0}
 \definecolor{WHITE}{gray}{1}
 \definecolor{RED}{rgb}{1,0,0}
 \definecolor{GREEN}{rgb}{0,1,0}
 \definecolor{BLUE}{rgb}{0,0,1}
 \definecolor{CYAN}{cmyk}{1,0,0,0}
 \definecolor{MAGENTA}{cmyk}{0,1,0,0}
 \definecolor{YELLOW}{cmyk}{0,0,1,0}
}

\usepackage{color}

\usepackage{ulem}

\usepackage{aecompl}%\usepackage[T1]{fontenc}

\usepackage{epsfig}\usepackage{dcolumn}\usepackage{bm}

\usepackage{babel}

\makeatother

\usepackage{babel}
\begin{document}

\title{Encrypting Majorana {F}ermions-\textit{qubits} as {B}ound {S}tates in the {C}ontinuum}

\author{L. H. Guessi$^{1,2}$, F. A. Dessotti$^{3}$, Y. Marques$^{3}$, L. S. Ricco$^{3}$,\\
G. M. Pereira$^{1}$, P. Menegasso$^{1,4}$, M. de Souza$^{1}$, and
A. C. Seridonio$^{1,3}$}

\affiliation{$^{1}$IGCE, Unesp - Univ Estadual Paulista, Departamento de F\'{i}sica,
13506-900, Rio Claro, SP, Brazil\\
 $^{2}$Instituto de Física de São Carlos, Universidade de São Paulo,
C.P. 369, São Carlos, SP, 13560-970, Brazil\\
 $^{3}$Departamento de F\'{i}sica e Qu\'{i}mica, Unesp - Univ Estadual
Paulista, 15385-000, Ilha Solteira, SP, Brazil\\
$^{4}$Instituto de Física ``Gleb Wataghin", Unicamp, Campinas-SP, 13083-859, Brazil}
\begin{abstract}
We theoretically investigate a topological Kitaev chain {connected to a double {quantum-dot (QD)}} setup hybridized with metallic leads. In this system, {we observe the emergence of two striking phenomena}: i) a decrypted {Majorana Fermion (MF)}-\textit{qubit} recorded over a single QD, {which is detectable by means of conductance measurements due} to the asymmetrical MF-leaked state into the QDs{;} ii) {an} encrypted \textit{qubit} {recorded} in both QDs when the leakage is symmetrical. In such a regime, we have {a cryptographylike manifestation}, {since} the MF-\textit{qubit} becomes bound states in the continuum, which {is not detectable} in conductance {experiments}.
\end{abstract}


\maketitle

\textit{Introduction.}---It is well known that Majorana fermions (MFs) zero-modes\cite{Alicea,Franz} are expected to appear bounded to the edges of a topological Kitaev chain\cite{Kitaev,Kitaev1,Kitaev2,Kitaev3,Kitaev4}. Interestingly enough, by approaching {the Kitaev} chain to a quantum dot (QD), the MF state leaks\cite{Vernek} into it and {manifests itself} as a zero-bias peak (ZBP) {in conductance measurements}. {The latter} reveals {experimentally} the  MF-\textit{qubit} recorded over the QD. {Indeed,} such a phenomenon was experimentally confirmed in a QD hybrid-nanowire made by InAs/Al\cite{wire2016} {with} huge spin-orbit {interaction} and magnetic fields, {being the nanowire placed close to} an \textit{s}-wave superconductor. It is worth mentioning that MFs can also emerge in the fractional quantum Hall state with filling{-}factor $\nu=5/2$\cite{QH}, in three-dimensional topological insulators\cite{TI}, at the core of superconducting vortices\cite{V1,V2,V3} and on the edges of {ferromagnetic atomic chains}  covering superconductors {with pronounced} spin-orbit interaction\cite{wire2,Jelena}{,} similarly to semiconducting nanowires\cite{wire1}. {In terms of technological applications}, MFs-\textit{qubits} {are of particular interest. This is because of} their topological protection against decoherence\cite{Kitaev}, {a key ingredient} for the achievement of efficient quantum computers.

\begin{figure}[!]
\includegraphics[width=0.48\textwidth,height=0.28\textheight]{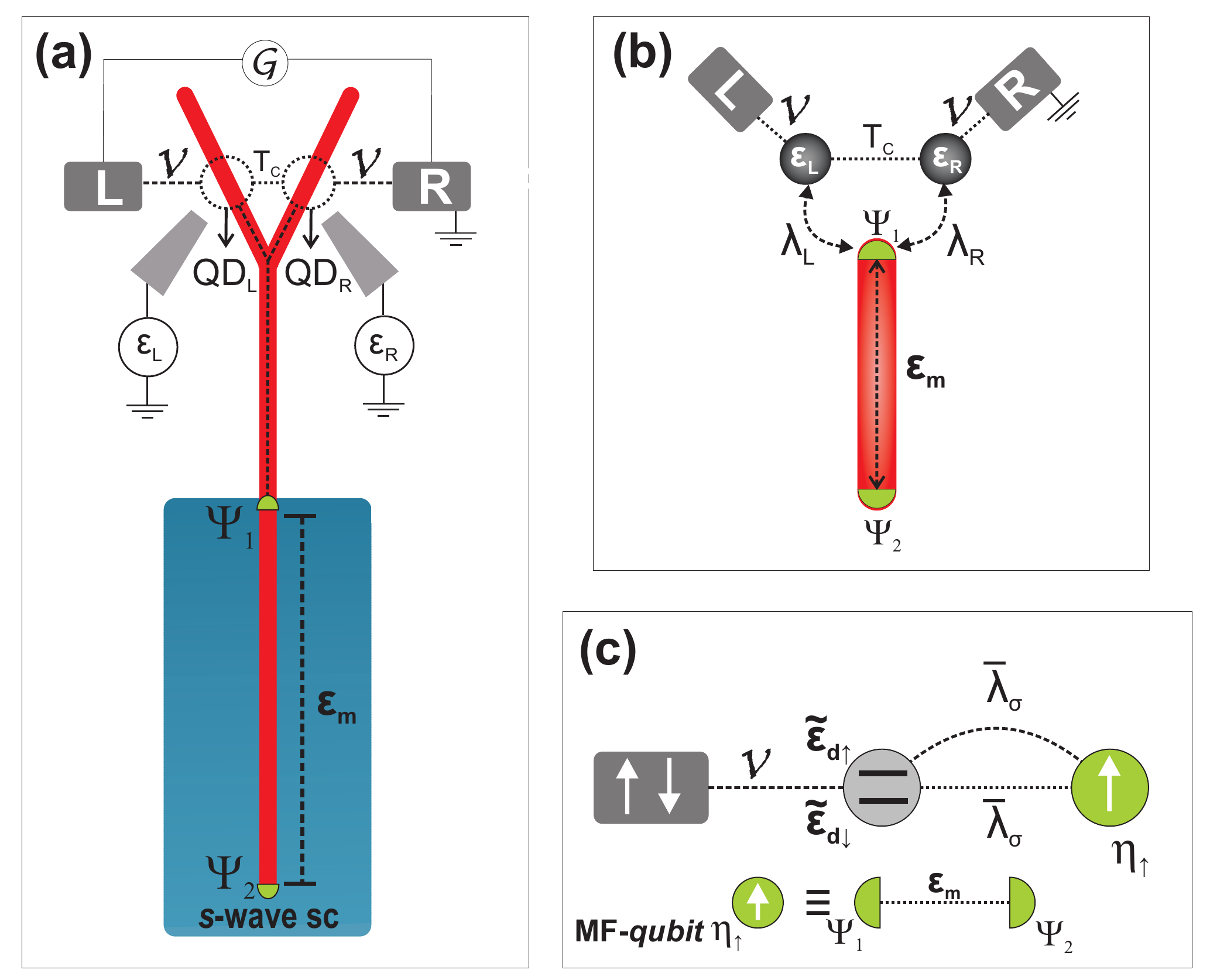}
\protect\protect\protect\protect\protect\protect\protect\protect\protect\protect\protect\protect\caption{\label{fig:Pic1}{(Color online) (a) Two QDs symmetrically coupled to leads via the hybridization $\mathcal{V}$ and asymmetrically to a topological Kitaev chain by means of the complex amplitudes $\lambda_L$ and $\lambda_R$. $\Psi_{1}$ and $\Psi_{2}$ are overlapped MFs with splitting energy $\varepsilon_{M}\to0$.} (b) Oversimplified sketch of panel (a). (c) Mapping of the original system from panel (b) onto the \textit{pseudospin} representation. The dressed \textit{pseudo}-Zeeman gap $\tilde{\epsilon}_{d\uparrow}-\tilde{\epsilon}_{d\downarrow}$ appears depicted within the central QD, which is side-coupled to the \textit{qubit} $\eta_{\uparrow},$ namely, the MF-\textit{qubit}. $\bar{\lambda}_{\sigma}$ identifies both the tunneling and the Cooper pair binding energy between the new QD and $\eta_{\uparrow}$, respectively given by the horizontal and semi-circular dashed lines.}
\end{figure}

{In this work, we {show} that the employment of two QDs, {as depicted in} Fig.\ref{fig:Pic1}(a), {enables} the cryptography of the MF-\textit{qubit} state $\eta_{\uparrow}=\frac{1}{\sqrt{2}}\left(\Psi_{1}+i\Psi_{2}\right)$ made by the MFs $\Psi_{1}$ and $\Psi_{2}$ with splitting energy $\varepsilon_{M}\to0$, where an encoded message can be written over these states of bits.} Our main theoretical findings rely on the interplay between the leakage effect and the so-called bound states in the continuum (BICs)\cite{Neuman,SH}.
{In this context, it is worth recalling} the underlying Physics {of such exotic excitations}. BICs were proposed by von Neumann and Wigner in $1929$\cite{Neuman} as quantum states with localized square-integrable wave functions, but surprisingly within the domain of the {energy} continuum region. Noteworthy, such states trap particles indefinitely.
{BICs constitute a current topic of} broad interest\cite{BICs}, appearing in several physical systems like graphene\cite{QPT,BIC1,Gong}, optics and photonics\cite{BIC2,BIC3,Exp1,Exp2}, arrangements exhibiting singular chirality\cite{Chiral} {and} Floquet-Hubbard states due to A.C. fields\cite{Floquet,AC}. Moreover, BICs assisted by MFs {enable} applications like the storage of \textit{qubits}\cite{Ricco} and the electrical current switch\cite{Dessotti} as well.
{It should be mentioned} that electrons trapped at BICs are prevented to decay into the energy continuum of the environment. Once BICs are undetectable by electrical conductance and accounting for the leakage effect, {we benefit} of such a remarkable invisibility feature of the BICs. {{Hence, for the} sake of simplicity, we {label} by cryptography of the MF-\textit{qubit} $\eta_{\uparrow}=\frac{1}{\sqrt{2}}\left(\Psi_{1}+i\Psi_{2}\right)$ when its ZBP signature disappears as a BIC, turning itself undetectable {via conductance measurements}.
{As it will be discussed below, we a}lso find an asymmetrical leakage of the MF-\textit{qubit}. In such a situation, the ZBP is visible {in} the conductance and we call {such a regime} by decrypted MF-\textit{qubit}, {since} the MF state from the Kitaev chain edge leaks solely into a single QD of the proposed setup (Fig.\ref{fig:Pic1}). Equivalently, the \textit{qubit} is recorded over this QD. Our decrypted MF case corresponds to the readout of the \textit{qubit} in QDs via charge measurement, i.e, the ZB-conductance, as proposed by Flensberg\cite{Flensberg}. {Otherwise}, the encrypted \textit{qubit} is achieved when the recording is symmetrical over the QDs, but with an invisible ZBP in the conductance in such a way that the readout is off, i.e., the decrypting is not allowed. In this regime, the MF-leaked state at zero-bias is split into the QDs, thus becoming BICs. Thereby, we propose that the switch on/off of the readout of the \textit{qubit} via the ZBP in the QDs consists in a manner of realizing quantum cryptography of the information written in the prepared MFs states $\Psi_{1}$ and $\Psi_2$.}

\textit{The Model.}---Below we describe theoretically the setup outlined in Fig.\ref{fig:Pic1}(a) with a topological Kitaev chain coupled to a double QD setup hybridized with metallic leads\cite{Baranger,Flensberg}. The oversimplified sketch of such a system is depicted in Fig.\ref{fig:Pic1}(b), which is ruled by the Hamiltonian
\begin{eqnarray}
\mathcal{H}_{\text{Full}} & = &\sum_{\alpha\mathbf{k}}\tilde{\varepsilon}_{\alpha\mathbf{k}}c_{\alpha\mathbf{k}}^{\dagger}c_{\alpha\mathbf{k}}+\sum_{\alpha}\varepsilon_{\alpha}d_{\alpha}^{\dagger}d_{\alpha}+T_{\text{{c}}}(d_{L}^{\dagger}d_{R}+\text{{H.c.}})\nonumber \\
 & + &\mathcal{V}\sum_{\alpha\mathbf{k}}(c_{\alpha\mathbf{k}}^{\dagger}d_{\alpha}+\text{{H.c.}})+\mathcal{H}_{\text{MFs}}, \label{eq:TIAM}
\end{eqnarray}
where the electrons in the lead $\alpha=L,R$ are described by the operator $c_{\alpha\mathbf{k}}^{\dagger}$ ($c_{\alpha\mathbf{k}}$) for the creation (annihilation) of an electron in a quantum state labeled by the wave number $\mathbf{k}$ and energy $\tilde{\varepsilon}_{\alpha\mathbf{k}}=\tilde{\varepsilon}_{\mathbf{k}}-\mu_{\alpha}$,
with $\mu_{\alpha}$ as the chemical potential. For the QDs coupled to the leads, $d_{\alpha}^{\dagger}$ ($d_{\alpha}$) creates (annihilates) an electron in the state $\varepsilon_{\alpha},$ which is gate tunable. The left-right QD coupling is $T_{\text{{c}}},$ while $\mathcal{V}$ stands for the hybridization between these QDs and the leads. {Concerning $\mathcal{H}_{\text{MFs}}$ we refer to Ref.[\onlinecite{Flensberg}] which accounts for QDs with large energy spacing of levels between spins up and down due to Zeeman splitting. Consequently, the spinless condition is fulfilled where only spin up state is relevant for the emerging topological superconductivity. In this way, our QDs do not depend on the charging energy as in Refs.[\onlinecite{wire2016},\onlinecite{Leijnse}], thus the QDs just couple asymmetrically to the Kitaev chain with complex tunneling amplitudes $\lambda_L$ and $\lambda_R$, respectively for the left and right QDs as follows
\begin{eqnarray}
\mathcal{H}_{\text{MFs}} & = & i\varepsilon_{M}\Psi_{1}\Psi_{2}+|\lambda_R|(e^{i\phi_R}d_{R}-e^{-i\phi_R}d_{R}^{\dagger})\Psi_{1}\nonumber \\
& + & |\lambda_L|(e^{i\phi_L}d_{L}-e^{-i\phi_L}d_{L}^{\dagger})\Psi_{1},\label{eq:HMBS}
\end{eqnarray}}
where $\Psi_{1}=\Psi_{1}^{\dagger}$ and $\Psi_{2}=\Psi_{2}^{\dagger}$
account for the MFs lying on the edges of the chain with overlap term $\varepsilon_{M}\sim e^{-L/\xi},$ wherein $L$ and $\xi$
designate respectively, the size of the Kitaev chain and the superconducting coherence length.

{We stress that, for a sake of simplicity, by employing the following substitutions $d_{L}=e^{-i\phi_L}[\left(\cos\theta\right)\tilde{d}_{\uparrow}-\left(\sin\theta\right)\tilde{d}_{\downarrow}]$,
$d_{R}=e^{-i\phi_R}[\left(\sin\theta\right)\tilde{d}_{\uparrow}+\left(\cos\theta\right)\tilde{d}_{\downarrow}]$,
$c_{\mathbf{k}L}=e^{-i\phi_L}[\left(\cos\theta\right)\tilde{c}_{\mathbf{k}\uparrow}-\left(\sin\theta\right)\tilde{c}_{\mathbf{k}\downarrow}]$
and $c_{\mathbf{k}R}=e^{-i\phi_R}[\left(\sin\theta\right)\tilde{c}_{\mathbf{k}\uparrow}+\left(\cos\theta\right)\tilde{c}_{\mathbf{k}\downarrow}]$
into the Hamiltonian of Eq.(\ref{eq:TIAM}), in particular at the zero-bias regime $(\mu_{\alpha}=0\equiv$Fermi level
of the leads$),$ we obtain}
\begin{eqnarray}
\mathcal{H}_{\text{Full}} & = & \sum_{\mathbf{k},\sigma}\tilde{\varepsilon}_{\mathbf{k}}\tilde{c}_{\mathbf{k}\sigma}^{\dagger}\tilde{c}_{\mathbf{k}\sigma}+\sum_{\sigma}\epsilon_{d\sigma}\tilde{d}_{\sigma}^{\dagger}\tilde{d}_{\sigma} + \mathcal{V}\sum_{\mathbf{k},\sigma}(\tilde{c}_{\mathbf{k}\sigma}^{\dagger}\tilde{d}_{\sigma}+\text{{H.c.}})\nonumber \\
& + & \mathcal{H}_{\text{MFs}},\label{eq:TIAM2}
\end{eqnarray}
which mimics an effective single QD coupled to an unique lead both
exhibiting an artificial spin degree of freedom $\sigma=\pm1$ $(\uparrow,\downarrow)$ (see Fig.\,\ref{fig:Pic1}(c) for such a representation).

We call attention that from now on, we label the aforementioned variable by \textit{pseudospin}. {As in Ref.[\onlinecite{Flensberg}], we have topological protection of our findings if the phase difference $\phi_L-\phi_R =2n\pi$ is fulfilled, with $n$ integer being tunable via magnetic flux, thus leading to $\cos(2\theta)=\frac{\triangle\epsilon \cos(\phi_L-\phi_R)}{\sqrt{4(T_{\text{{c}}})^{2}+(\triangle\epsilon)^{2}}}$,}
$\triangle\epsilon=\varepsilon_{L}-\varepsilon_{R}$ as the detuning
of the original spinless QDs, the \textit{pseudo}-Zeeman gap $\epsilon_{d\uparrow}-\epsilon_{d\downarrow},$
with $\epsilon_{d\sigma}=\frac{(\varepsilon_{L}+\varepsilon_{R})}{2}+\frac{\sigma}{2}\sqrt{4(T_{\text{{c}}})^{2}+(\triangle\epsilon)^{2}}$
and
\begin{align}
\mathcal{H}_{\text{MFs}} & =\varepsilon_{M}(\eta^{\dagger}_{\uparrow}\eta_{\uparrow}-\frac{1}{2})+\sum_{\sigma}\bar{\lambda}_{\sigma}(\tilde{d}_{\sigma}\eta^{\dagger}_{\uparrow}+\tilde{d}_{\sigma}\eta_{\uparrow}+\text{{H.c.}}),\label{eq:HMBS2}
\end{align}
where we have used $\Psi_{1}=\frac{1}{\sqrt{2}}\left(\eta^{\dagger}_{\uparrow}+\eta_{\uparrow}\right)$ and $\Psi_{2}=\frac{i}{\sqrt{2}}\left(\eta^{\dagger}_{\uparrow}-\eta_{\uparrow}\right)$
in order to build the \textit{qubit} $\eta_{\uparrow}$ composed by the MFs, namely the MF-\textit{qubit}, {with $\bar{\lambda}_{\uparrow}=\frac{1}{\sqrt{2}}(|\lambda_L|\cos\theta+|\lambda_R|\sin\theta)$ and $\bar{\lambda}_{\downarrow}=\frac{1}{\sqrt{2}}(|\lambda_R|\cos\theta-|\lambda_L|\sin\theta)$
as \textit{pseudospin}-dependent amplitudes.} As a result, the \textit{pseudo}-Zeeman gap becomes dressed by such an interaction, i.e., $\tilde{\epsilon}_{d\uparrow}-\tilde{\epsilon}_{d\downarrow},$ which will be addressed later on.

{We call attention to the system Hamiltonian mapping into Eq.(\ref{eq:HMBS2}), where one can recognize that the device of Fig.\,\ref{fig:Pic1}(b) is equivalent to the QD $\tilde{d}_{\sigma}$ emulating the two original spinless left and right QDs, in particular side-coupled to $\eta_{\uparrow}$, which corresponds to a QD replacing the Kitaev chain. This opens the possibility of reproducing experimentally the same phenomenon reported here for the topological Kitaev chain by employing QDs, but in the presence of a delocalized Cooper pair split into $\tilde{d}_{\sigma}$ and $\eta_{\uparrow}$ with pairing amplitude $\bar{\lambda}_{\sigma}$ as the terms $\bar{\lambda}_{\sigma}(\tilde{d}_{\sigma}\eta_{\uparrow}+\text{H.c.})$ point out. Besides, the normal tunneling between these QDs should be also equal to $\bar{\lambda}_{\sigma}$, i.e., $\bar{\lambda}_{\sigma}(\tilde{d}_{\sigma}\eta^{\dagger}_{\uparrow}+\text{H.c.})$, just in order to ensure the emergence of MFs at the so-called ``sweet spot" as predicted in Ref.[\onlinecite{Flensberg2}] by Flensberg. In such a work, the equivalence of the topological Kitaev chain with a QD system is established by means of an analogous Hamiltonian to our Eq.(\ref{eq:HMBS2}). In this way, this system of QDs hosting MFs becomes an experimental alternative with respect to the topological Kitaev chain. Noteworthy, this QD-like alternative system with MFs was already explored by some of us in Ref.\cite{Ricco1} within the context of adatoms and STM tips as well as the case of a zero-mode from a regular normal side-coupled QD to a central QD region\cite{Ricco2}. For this latter, the \textit{qubit} $\eta_{\uparrow}$ without the Cooper pairing amplitude (proximity effect) when encrypted would be still protected against the decoherence of the surroundings due to the BIC nature of the state which decouples it from the environment, thus preventing a finite conductance through this channel. Equivalently, BICs do not depend on the proximity effect to occur. However, the decrypted \textit{qubit} case would not be protected in the same way, once it couples to the environment in contrast to a MF-\textit{qubit}, which is topologically protected characterized by a pinned ZBP. This characteristic plays the main difference from a regular fermionic zero-mode, wherein the expected ZBP is destroyed by changing external parameters as outlined in Fig.4(a) of Ref.\cite{Ricco2} and the readout of the \textit{qubit} in the central QD region is compromised as a result.}

In what follows, we use the Landauer-Büttiker formula for the zero-bias
conductance $\text{\ensuremath{\mathcal{G}}}$\cite{book} to
analyze the transport through the QDs, which is
\begin{equation}
\text{\ensuremath{\mathcal{G}}}=\frac{e^{2}}{h}\int d\varepsilon\left(-\frac{\partial f_{F}}{\partial\varepsilon}\right)\text{\ensuremath{\mathcal{T}}}_{\text{{Total}}},\label{eq:G}
\end{equation}
where $f_{F}$ stands for the Fermi-Dirac distribution, $\text{\ensuremath{\mathcal{T}}}_{\text{{Total}}}=\sum_{j}\text{\ensuremath{\mathcal{T}}}_{jj}+\sum_{j}\text{\ensuremath{\mathcal{T}}}_{j\bar{j}}$
encodes the system total transmittance with $j=L,R$ respectively
for $\bar{j}=R,L$ to correlate distinct QDs, in which $\mathcal{T}_{jl}=\mathcal{T}_{\uparrow\uparrow}+\mathcal{T}_{\downarrow\downarrow}+\mathcal{T}_{\uparrow\downarrow}+\mathcal{T}_{\downarrow\uparrow}$
dictates the transmittance through the channels $l,j=L,R$ in terms of the coefficients $\mathcal{T}_{\sigma\tilde{\sigma}}$ for the \textit{pseudospin} representation.

Furthermore, $\mathcal{T}_{jl}=\pi\Gamma\rho_{jl}$ depends upon the Anderson broadening $\Gamma=\pi\mathcal{V}^{2}\sum_{\mathbf{k}}\delta(\varepsilon-\tilde{\varepsilon}_{\mathbf{k}})$
\cite{Anderson} and $\rho_{jl}=(-1/\pi)\text{{Im}}({\tilde{\mathcal{G}}_{{d}_{j},{d}_{l}}}),$ the densities of states for the spinless QDs from the Hamiltonian of Eq.(\ref{eq:TIAM}) in terms of the retarded Green's functions $\tilde{\mathcal{G}}_{{d}_{j},{d}_{l}}$, which
are given by
\begin{align}
\rho_{LL} & =-\frac{1}{\pi}\text{{Im}}\{\cos^{2}\theta\tilde{\mathcal{G}}_{\tilde{d}_{\uparrow},\tilde{d}_{\uparrow}}+\sin^{2}\theta\tilde{\mathcal{G}}_{\tilde{d}_{\downarrow},\tilde{d}_{\downarrow}}\nonumber \\
  &-\sin\theta\cos\theta(\tilde{\mathcal{G}}_{\tilde{d}_{\downarrow},\tilde{d}_{\uparrow}}+\tilde{\mathcal{G}}_{\tilde{d}_{\uparrow},\tilde{d}_{\downarrow}})\},\label{eq:rhoLL}
\end{align}
\begin{align}
\rho_{RR} & =-\frac{1}{\pi}\text{{Im}}\{\sin^{2}\theta\tilde{\mathcal{G}}_{\tilde{d}_{\uparrow},\tilde{d}_{\uparrow}}+\cos^{2}\theta\tilde{\mathcal{G}}_{\tilde{d}_{\downarrow},\tilde{d}_{\downarrow}}\nonumber \\
  &+\sin\theta\cos\theta(\tilde{\mathcal{G}}_{\tilde{d}_{\downarrow},\tilde{d}_{\uparrow}}+\tilde{\mathcal{G}}_{\tilde{d}_{\uparrow},\tilde{d}_{\downarrow}})\},\label{eq:rhoRR}
\end{align}
\begin{align}
\rho_{RL} & =-\frac{1}{\pi}\text{{Im}}\{\sin\theta\cos\theta(\tilde{\mathcal{G}}_{\tilde{d}_{\uparrow},\tilde{d}_{\uparrow}}-\tilde{\mathcal{G}}_{\tilde{d}_{\downarrow},\tilde{d}_{\downarrow}})\nonumber \\
  &+\cos^{2}\theta\tilde{\mathcal{G}}_{\tilde{d}_{\downarrow},\tilde{d}_{\uparrow}}-\sin^{2}\theta\tilde{\mathcal{G}}_{\tilde{d}_{\uparrow},\tilde{d}_{\downarrow}}\}\label{eq:rhoRL}
\end{align}
and
\begin{align}
\rho_{LR} & =-\frac{1}{\pi}\text{{Im}}\{\sin\theta\cos\theta(\tilde{\mathcal{G}}_{\tilde{d}_{\uparrow},\tilde{d}_{\uparrow}}-\tilde{\mathcal{G}}_{\tilde{d}_{\downarrow},\tilde{d}_{\downarrow}})\nonumber \\
  &-\sin^{2}\theta\tilde{\mathcal{G}}_{\tilde{d}_{\downarrow},\tilde{d}_{\uparrow}}+\cos^{2}\theta\tilde{\mathcal{G}}_{\tilde{d}_{\uparrow},\tilde{d}_{\downarrow}}\},\label{eq:rhoLR}
\end{align}
here written as functions of the retarded Green's functions $\tilde{\text{\ensuremath{\mathcal{G}}}}_{\tilde{d}_{\sigma},\tilde{d}_{\tilde{\sigma}}}$
within the mapping on the \textit{pseudospin} degree. To evaluate $\tilde{\text{\ensuremath{\mathcal{G}}}}_{\tilde{d}_{\sigma},\tilde{d}_{\tilde{\sigma}}},$ we should employ the equation-of-motion method\cite{book} by using Eqs.(\ref{eq:TIAM2}) and (\ref{eq:HMBS2}) as follows: $\varepsilon\tilde{\mathcal{G}}_{\tilde{d}_{\sigma},\tilde{d}_{\tilde{\sigma}}}=[\tilde{d}_{\sigma},\tilde{d}_{\tilde{\sigma}}^{\dagger}]_{+}+\tilde{\mathcal{G}}_{\left[\tilde{d}_{\sigma},\mathcal{\mathcal{H_{\text{Full}}}}\right],\tilde{d}_{\tilde{\sigma}}}.$
As a result, we find the linear system:
\begin{align}
(\varepsilon-\epsilon_{d\sigma}-\bar{\lambda}_{\sigma}^{2}K+i\Gamma)\tilde{\mathcal{G}}_{\tilde{d}_{\sigma},\tilde{d}_{\sigma}}-\bar{\lambda}_{\bar{\sigma}}\bar{\lambda}_{\sigma}K\tilde{\mathcal{G}}_{\tilde{d}_{\bar{\sigma}},\tilde{d}_{\sigma}}\nonumber \\
+\bar{\lambda}_{\sigma}^{2}K\tilde{\mathcal{G}}_{\tilde{d}_{\sigma}^{\dagger},\tilde{d}_{\sigma}}
+\bar{\lambda}_{\sigma}\bar{\lambda}_{\bar{\sigma}}K\tilde{\mathcal{G}}_{\tilde{d}_{\bar{\sigma}}^{\dagger},\tilde{d}_{\sigma}}=1,\label{eq:G1}
\end{align}
\begin{align}
-\bar{\lambda}_{\sigma}\bar{\lambda}_{\bar{\sigma}}K\tilde{\mathcal{G}}_{\tilde{d}_{\bar{\sigma}},\tilde{d}_{\bar{\sigma}}}+(\varepsilon-\epsilon_{d\sigma}-\bar{\lambda}_{\sigma}^{2}K+i\Gamma)\tilde{\mathcal{G}}_{\tilde{d}_{\sigma},\tilde{d}_{\bar{\sigma}}} \nonumber \\
+\bar{\lambda}_{\sigma}\bar{\lambda}_{\bar{\sigma}}K\tilde{\mathcal{G}}_{\tilde{d}_{\bar{\sigma}}^{\dagger},\tilde{d}_{\bar{\sigma}}}+\bar{\lambda}_{\sigma}^{2}K\tilde{\mathcal{G}}_{\tilde{d}_{\sigma}^{\dagger},\tilde{d}_{\bar{\sigma}}}=0,\label{eq:G2}
\end{align}
\begin{align}
\bar{\lambda}_{\sigma}^{2}K\tilde{\mathcal{G}}_{\tilde{d}_{\sigma},\tilde{d}_{\sigma}}+\bar{\lambda}_{\sigma}\bar{\lambda}_{\bar{\sigma}}K\tilde{\mathcal{G}}_{\tilde{d}_{\bar{\sigma}},\tilde{d}_{\sigma}} \nonumber \\
+(\varepsilon+\epsilon_{d\sigma}-\bar{\lambda}_{\sigma}^{2}K+i\Gamma)\tilde{\mathcal{G}}_{\tilde{d}_{\sigma}^{\dagger},\tilde{d}_{\sigma}}-\bar{\lambda}_{\sigma}\bar{\lambda}_{\bar{\sigma}}K\tilde{\mathcal{G}}_{\tilde{d}_{\bar{\sigma}}^{\dagger},\tilde{d}_{\sigma}}=0\label{eq:G4}
\end{align}
and
\begin{align}
\bar{\lambda}_{\sigma}\bar{\lambda}_{\bar{\sigma}}K\tilde{\mathcal{G}}_{\tilde{d}_{\bar{\sigma}},\tilde{d}_{\bar{\sigma}}}+\bar{\lambda}_{\sigma}^{2}K\tilde{\mathcal{G}}_{\tilde{d}_{\sigma},\tilde{d}_{\bar{\sigma}}}
-\bar{\lambda}_{\bar{\sigma}}\bar{\lambda}_{\sigma}K\widetilde{\mathcal{G}}_{\tilde{d}_{\bar{\sigma}}^{\dagger},\tilde{d}_{\bar{\sigma}}}\nonumber \\
+(\varepsilon+\epsilon_{d\sigma}-\bar{\lambda}_{\sigma}^{2}K+i\Gamma)\tilde{\mathcal{G}}_{\tilde{d}_{\sigma}^{\dagger},\tilde{d}_{\bar{\sigma}}}=0 & ,\label{eq:G5}
\end{align}
where $\bar{\sigma}$ is the opposite of $\sigma$ and $K=(\varepsilon+\varepsilon_{M})^{-1}+(\varepsilon-\varepsilon_{M})^{-1}.$ To perform the analysis of the model in the next section, we make explicit that we have solved the current system numerically.

\begin{figure}[!]
\centering
\includegraphics[width=0.48\textwidth,height=0.30\textheight]{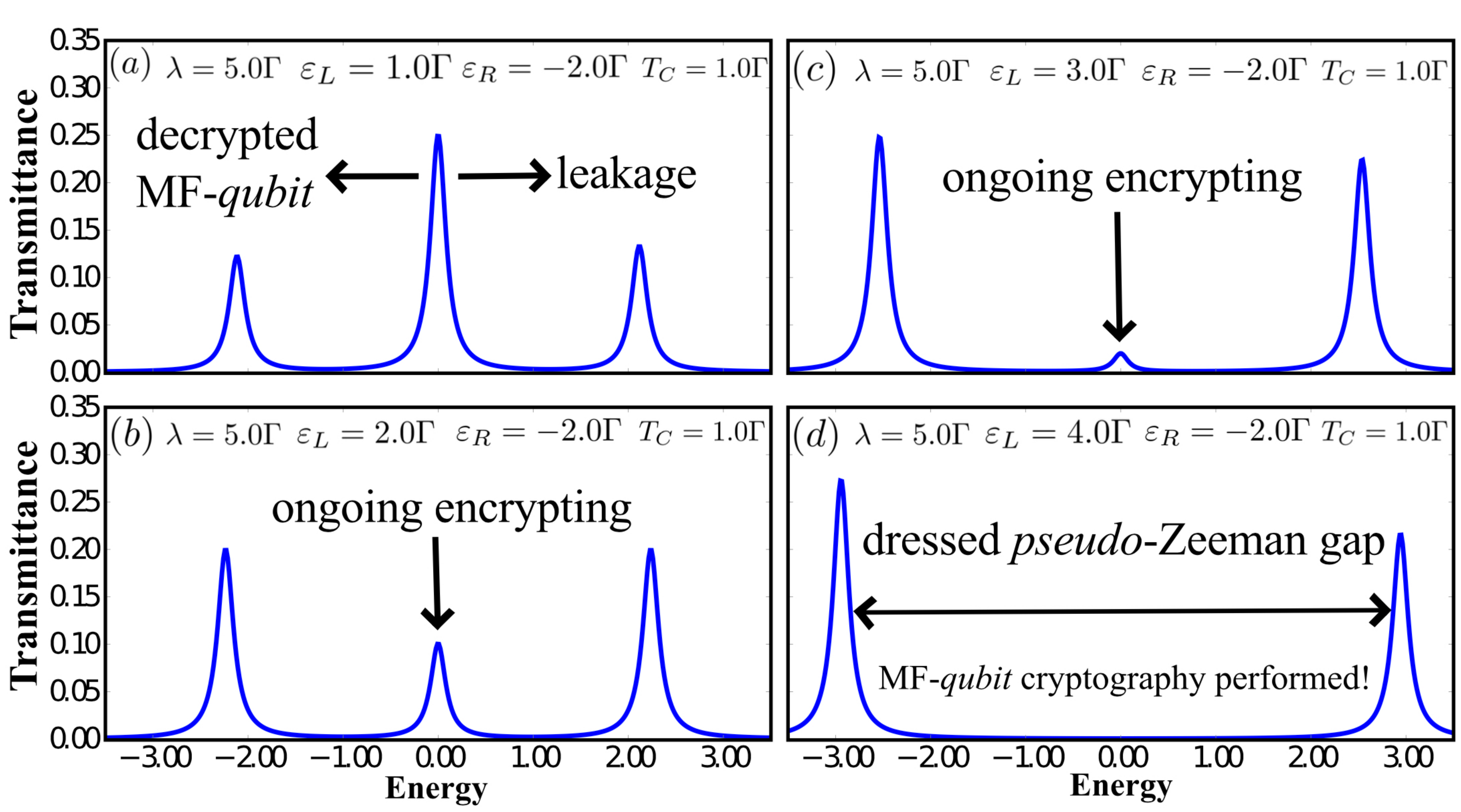} \protect\caption{\label{fig:Pic2}(Color online) $\text{\ensuremath{\mathcal{T}}}_{\text{{Total}}}$
as a function of $\varepsilon$: (a) The ZBP gives the asymmetrical leakage of the MF-\textit{qubit} $\eta_{\uparrow}$ into the left QD (see also Fig.\ref{fig:Pic3}). (b)-(c) The increasing of $\varepsilon_{R}$ yields the process for encrypting this \textit{qubit}, which is characterized by the quenching of the ZBP amplitude. (d) Here the ZBP (the MF-\textit{qubit}) is hidden as BICs equally split into the QDs, where only the dressed \textit{pseudo}-Zeeman gap is visible (see also Fig.\ref{fig:Pic4}).}
\end{figure}

\begin{figure}[!]
\centering
\includegraphics[width=0.48\textwidth,height=0.60\textheight]{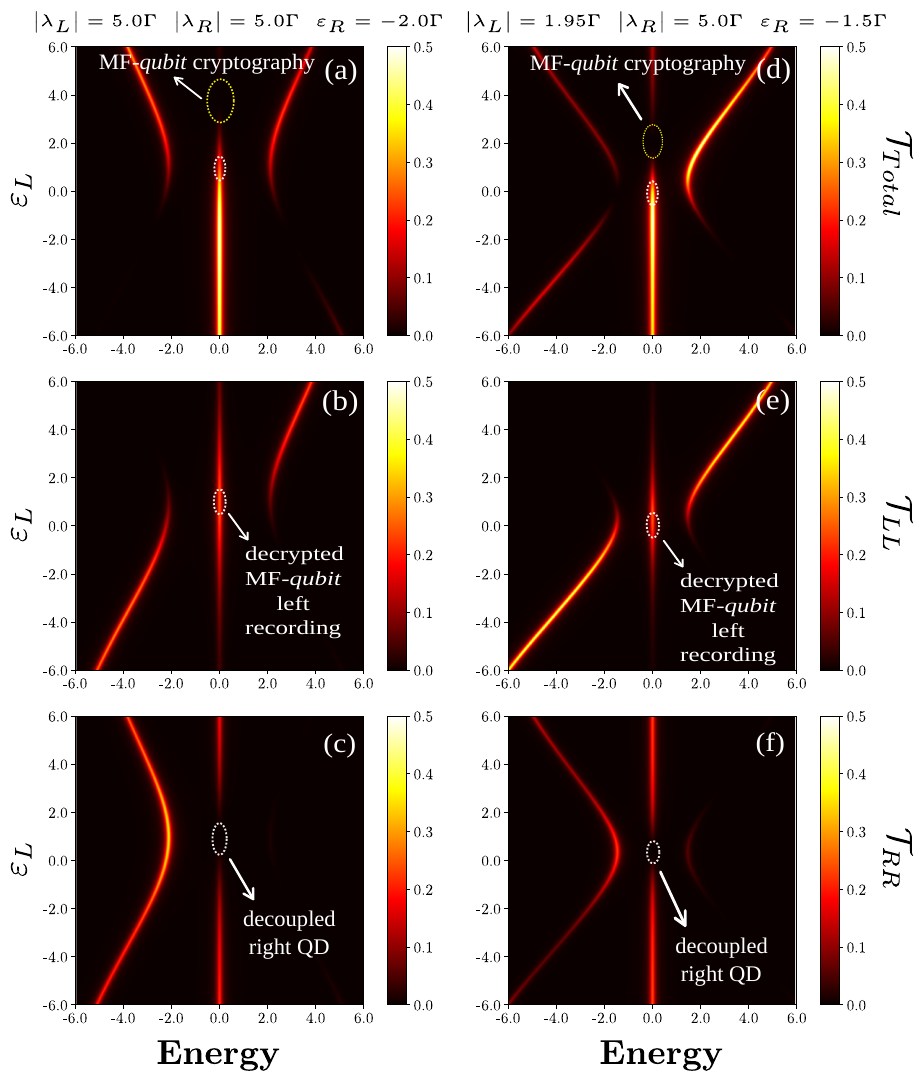}
\protect\protect\protect\protect\protect\protect\protect\protect\protect\protect\protect\protect
\caption{\label{fig:Pic3}(Color online) Density plots of: (a) $\text{\ensuremath{\mathcal{T}}}_{\text{{Total}}}$,
(b) $\mathcal{T}_{LL}$ and (c) $\mathcal{T}_{RR}$ spanned by $\varepsilon_{L}$ and $\varepsilon,$ with $\varepsilon_{R}=-2\Gamma,$
{$|\lambda_L|=|\lambda_R|=\lambda=5\Gamma$} and $T_{\text{{c}}}=1\Gamma.$ The ellipses depicted show at zero-bias: i) the region for the MF-\textit{qubit }cryptography (yellow dashed ellipse) in (a) and ii) the corresponding for the decrypted MF-\textit{qubit} left recording (white dashed
ellipse) in (a) and (b), due to the right QD entirely decoupled from the system as panel (c) shows (white dashed
ellipse). {Panels (d),(e) and (f) for $|\lambda_L|\neq|\lambda_R|$ give qualitatively the same of (a),(b) and (c) thus ensuring the topological robustness of the results, i.e., the BICs (the encrypted MF-\textit{qubit}) and the decrypted MF-\textit{qubit} left recording still occur, but for different set of parameters.}}
\end{figure}

\begin{figure}[!]
\centering
\includegraphics[width=0.48\textwidth,height=0.25\textheight]{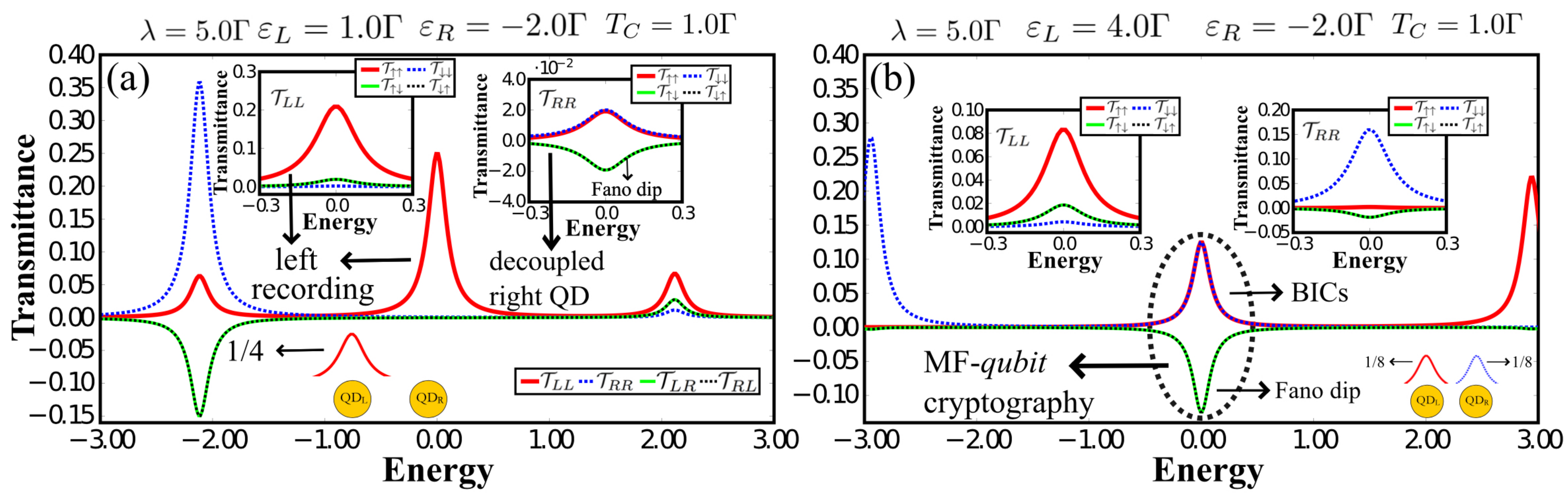}
\protect\protect\protect\protect\protect\protect\protect\protect\protect\protect\protect\protect\caption{\label{fig:Pic4}(Color online) $\mathcal{T}_{jl}$
in (a) characterizing the decrypted MF-\textit{qubit}
left recording. $\mathcal{T}_{LL}$ shows a ZBP with amplitude $1/4$,
while $\mathcal{T}_{RR}$ does not: the inset reveals that
$\mathcal{T}_{RR}$ exhibits $\mathcal{T}_{\uparrow\uparrow}+\mathcal{T}_{\downarrow\downarrow}$
perfectly phase shifted by $\pi$ with respect to \textit{$\mathcal{T}_{\uparrow\downarrow}+\mathcal{T}_{\downarrow\uparrow}$
}(Fano dip). As aftermath, this QD is disconnected from the system.
In (b), we have the MF-\textit{qubit} cryptography: in $\mathcal{T}_{RR},$ the Fano
dip is not perfect as before. However, a Fano dip in $\mathcal{T}_{LR}+\mathcal{T}_{RL}$
interferes destructively and exactly with $\mathcal{T}_{LL}+\mathcal{T}_{RR}.$
It means that MF-\textit{qubit} is hidden as BICs equally divided into the QDs.}
\end{figure}

\textit{Results and Discussion.}---In the simulations below the temperature $T=0$ is assumed and $\Gamma=40\mu eV$\cite{Vernek,Anderson} as the energy scale. The topological Kitaev chain, for a sake of simplicity, is treated as very large, which imposes $\varepsilon_{M}\to0.$ Thus, in order to make explicit the phenomenon of MF-\textit{qubit }cryptography, we begin discussing the picture requested for the emergence of such in Fig.\ref{fig:Pic2}. Fig.\ref{fig:Pic2}(a) accounts for $\varepsilon_{R}=-2\Gamma,$ {$|\lambda_L|=|\lambda_R|=\lambda=5\Gamma,$} $T_{\text{{c}}}=1\Gamma$ and $\varepsilon_{L}=1\Gamma,$ where we verify a ZBP with amplitude of $1/4$ in $\text{\ensuremath{\mathcal{T}}}_{\text{{Total}}}$ of Eq.(\ref{eq:G}) as a function of $\varepsilon.$ This detectable resonance represents the leakage of the MF-\textit{qubit} $\eta_{\uparrow}$ into the double QD setup. Additionally, it also encodes the recording of a decrypted MF-\textit{qubit} over the left QD, which will be elucidated later on via Figs.\ref{fig:Pic3} and \ref{fig:Pic4}. On this ground, let us consider the sequence of panels from (b) to (d), which describes the \textit{qubit }cryptography itself: by changing just $\varepsilon_{L},$ we notice that the ZBP amplitude becomes reduced progressively up to entire quenching in Fig.\ref{fig:Pic2}(d). In this case, solely a couple of peaks stay visible denoting the dressed \textit{pseudo}-Zeeman gap $\tilde{\epsilon}_{d\uparrow}-\tilde{\epsilon}_{d\downarrow}.$ Indeed, we will clarify that the ZBP becomes BICs, being undetectable by  $\text{\ensuremath{\mathcal{T}}}_{\text{{Total}}}.$ It means that if the ZBP is not perceived, we have the accomplishment of the MF-\textit{qubit} cryptography, which appears addressed in detail by Figs.\ref{fig:Pic3} and \ref{fig:Pic4}.

{Fig.\ref{fig:Pic3} exhibits the density plots for $\text{\ensuremath{\mathcal{T}}}_{\text{{Total}}},$$\text{\ensuremath{\mathcal{T}}}_{LL}$
and $\text{\ensuremath{\mathcal{T}}}_{RR}$ spanned by the axis $\varepsilon$
and $\varepsilon_{L}$ for fixed $\varepsilon_{R}=-2\Gamma$ ($\varepsilon_{R}=-1.5\Gamma$) $\lambda=5\Gamma$ ($|\lambda_L|=1.95\Gamma$ and $|\lambda_R|=5\Gamma$)
and $T_{\text{{c}}}=1\Gamma.$} It is worth noticing that all panels
in {Figs.\ref{fig:Pic3}(a)-(f)} present a ZBP structure.
However, each one reveals different aspects on the leakage effect. For instance, in
{Fig.\ref{fig:Pic3}(a)((d))} we highlight the upper region marked by
a yellow dashed ellipse: it gives the domain where the MF-\textit{qubit
}cryptography is allowed, once the ZBP is absent. {Figs.\ref{fig:Pic3}(b)((e)) and (c)((f))} contain the asymmetrical leakage into
the QDs and the decrypted MF-\textit{qubit} left recording as well. {Notice that in the latter,
nearby $\varepsilon_{L}=1\Gamma$ ($\varepsilon_{L}=0$), the right QD decouples from the setup, due to $\text{\ensuremath{\mathcal{T}}}_{RR}=0.$}
This region is then identified by white dashed
ellipses {in panels (a)-(f) of the same figure.} As a result, the MF state is recorded solely at the left
QD as {Fig.\ref{fig:Pic3}(b)((e))} ensures. {This corresponds to the readout of the \textit{qubit} by a charge measurement as proposed by Flensberg\cite{Flensberg}. Notice that both $\text{\ensuremath{\mathcal{T}}}_{LL}$ and $\text{\ensuremath{\mathcal{T}}}_{RR}$ share the same brightness in their scales, thus pointing out the symmetrical leakage of the MF zero-mode is robust against asymmetrical couplings.} Concerning the satellite
arcs aside the ZBP in {Figs.\ref{fig:Pic3}(a)-(f)}, they account for the dressed \textit{pseudo}-Zeeman gap $\tilde{\epsilon}_{d\uparrow}-\tilde{\epsilon}_{d\downarrow}.$ These arcs are predominantly absent, as we can see, at the lower region of Fig.\ref{fig:Pic3}(a). This points out that BICs away from the ZB limit are also reliable in this device. {Thereby, in order to fully understand the underlying physics on the decrypted
MF-\textit{qubit} left recording versus the MF-\textit{qubit }cryptography,
we should consider Fig.\ref{fig:Pic4}, which uses the same parameters of Fig.\ref{fig:Pic2} just for a matter of choice, once for the emergence of the BICs the leakage is always symmetrical even with asymmetrical couplings $\lambda_L$ and $\lambda_R$ as Figs.\ref{fig:Pic3}(e) and (f) ensure.}

In Fig.\ref{fig:Pic4}(a) the analysis of $\mathcal{T}_{jl}$ shows that the leakage of the MF occurs only over the left QD. In this way, the decrypted MF-\textit{qubit} situation is achieved: $\mathcal{T}_{LL}$ exhibits a ZBP with amplitude $1/4$ in contrast to $\mathcal{T}_{RR}.$
Thus in order to understand such an issue, we should focus on the insets. $\mathcal{T}_{RR}$ presents $\mathcal{T}_{\uparrow\uparrow}+\mathcal{T}_{\downarrow\downarrow}$
perfectly phase shifted by $\pi$ with respect to \textit{$\mathcal{T}_{\uparrow\downarrow}+\mathcal{T}_{\downarrow\uparrow}$}(Fano
dip)\cite{Fano1,Fano2}, thus resulting in a decoupled QD from the setup. For $\mathcal{T}_{LL},$
$\mathcal{T}_{\uparrow\uparrow}+\mathcal{T}_{\downarrow\downarrow}$
and \textit{$\mathcal{T}_{\uparrow\downarrow}+\mathcal{T}_{\downarrow\uparrow}$
}interfere constructively. In Fig.\ref{fig:Pic4}(b) for $\mathcal{T}_{RR},$ the Fano dip\textit{ }in\textit{
$\mathcal{T}_{\uparrow\downarrow}+\mathcal{T}_{\downarrow\uparrow}$
}is not perfect as previously and does not cancel \textit{ }$\mathcal{T}_{\uparrow\uparrow}+\mathcal{T}_{\downarrow\downarrow}$
anymore. Particularly, the Fano dip found in $\mathcal{T}_{LR}+\mathcal{T}_{RL}$
interferes destructively and perfectly with the peak in $\mathcal{T}_{LL}+\mathcal{T}_{RR}.$ Finally, this yields the MF-\textit{qubit} cryptography here proposed.  In this way, the recording of the \textit{qubit} is found secure at two apart sites and hidden as BICs, which are equally split into the QDs and with amplitude
$1/8$ each. These processes appear outlined in the sketches placed at the lower region of Figs.\ref{fig:Pic4}(a) and (b).

Interestingly enough, the underlying physics of this cryptography assisted by BICs has a simple picture: the electronic waves traveling forth and back between the QDs ($\mathcal{T}_{LR}+\mathcal{T}_{RL}$), in particular at zero-bias, interfere destructively with those waves that only pass through these QDs ($\mathcal{T}_{LL}+\mathcal{T}_{RR}$) and as a result, the BICs within the latter emerge. Regarding the satellite arcs aside the ZBP in Figs.\ref{fig:Pic4}(a) and (b), we should mention that they are also the result of interference processes in $\mathcal{T}_{jl}$ as observed.

\textit{Conclusions.}---In summary, we have found theoretically that the cryptography of the MF-\textit{qubit} is feasible in the system of Fig.\ref{fig:Pic1}(a). We have showed that the recording of the MF-\textit{qubit} over a single QD is due to an asymmetrical leakage of the MF state into the QDs. The encrypted MF-\textit{qubit} is performed when the leaking is symmetrical, wherein the MF-leaked state becomes BICs. {Thus we reveal a switch on/off mechanism for the readout of the \textit{qubit} $\eta_{\uparrow}=\frac{1}{\sqrt{2}}\left(\Psi_{1}+i\Psi_{2}\right)$ by means of its ZBP fingerprint on the QDs, which provides a way of performing quantum cryptography regarding the message written inside the MFs states $\Psi_{1}$ and $\Psi_2$ initially prepared at the edges of a topological Kitaev chain. Therefore, we trust that our findings can be applied to quantum processing issues in topological quantum computation devices.}

\textit{Acknowledgments.}---This work was supported by CNPq (307573/2015-0), CAPES and S{ã}o Paulo Research Foundation (FAPESP): grants 2015/26655-9 and 2015/23539-8. We thank L.N. Oliveira for helpful suggestions.

\end{document}